\documentclass[12pt]{iopart}
\pdfoutput=1

\usepackage{iopams}

\usepackage{cite}
\usepackage{url}
\usepackage{color}
\usepackage{graphicx}
\usepackage{epstopdf}

\usepackage{amssymb}
\usepackage{color}

\begin{document}


\title[]{Seeking for a fingerprint: analysis of point processes in actigraphy recording}
\author{Ewa Gudowska-Nowak$^*$}
\address{Marian Smoluchowski Institute of Physics and Mark Kac Center for Complex Systems Research, Jagiellonian University, ul. {\L}ojasiewicza 11, 30--348 Krak\'ow, Poland}
\address{Malopolska Center of Biotechnology, Jagiellonian University, ul. Gronostajowa 7, 30---348 Krak\'ow, Poland}
\address{$^*$ Contributing author: ewa.gudowska-nowak@uj.edu.pl}
\author{Jeremi K. Ochab}
\address{Marian Smoluchowski Institute of Physics and Mark Kac Center for Complex Systems Research, ul. {\L}ojasiewicza 11, 30--348 Krak\'ow, Poland}
\author{Katarzyna Ole\'s}
\address{Marian Smoluchowski Institute of Physics and Mark Kac Center for Complex Systems Research,  ul. {\L}ojasiewicza 11, 30--348 Krak\'ow, Poland}
\author{Ewa Beldzik}
\address{Malopolska Center of Biotechnology, Jagiellonian University, ul. Gronostajowa 7, 30---348 Krak\'ow, Poland}
\author{Dante R. Chialvo}
\address{CONICET, Buenos Aires, Argentina}
\author{Aleksandra Domagalik}
\address{Malopolska Center of Biotechnology, Jagiellonian University, ul. Gronostajowa 7, 30---348 Krak\'ow, Poland}
\author{Magdalena F{\c{a}}frowicz}
\address{Malopolska Center of Biotechnology, Jagiellonian University, ul. Gronostajowa 7, 30---348 Krak\'ow, Poland}
\author{Tadeusz Marek}
\address{Malopolska Center of Biotechnology, Jagiellonian University, ul. Gronostajowa 7, 30---348 Krak\'ow, Poland}
\author{Maciej A. Nowak}
\address{Marian Smoluchowski Institute of Physics and Mark Kac Center for Complex Systems Research, Jagiellonian University, ul. {\L}ojasiewicza 11, 30--348 Krak\'ow, Poland }
\author{Halszka Ogi\'nska}
\address{Malopolska Center of Biotechnology, Jagiellonian University, ul. Gronostajowa 7, 30---348 Krak\'ow, Poland}
\author{Jerzy Szwed}
\address{Marian Smoluchowski Institute of Physics and Mark Kac Center for Complex Systems Research, Jagiellonian University, ul. {\L}ojasiewicza 11, 30--348 Krak\'ow, Poland }
\author{Jacek Tyburczyk}
\address{Marian Smoluchowski Institute of Physics and Mark Kac Center for Complex Systems Research, Jagiellonian University, ul. {\L}ojasiewicza 11, 30--348 Krak\'ow, Poland }
\begin{abstract}

Motor activity of humans displays complex temporal fluctuations which can be characterized by scale-invariant statistics, thus documenting that structure and fluctuations of such kinetics remain similar over a broad range of time scales. Former studies on humans regularly deprived of sleep or suffering from sleep disorders predicted change in the invariant scale parameters with respect to those representative for healthy subjects.
In this study we investigate the signal patterns from actigraphy recordings by means of characteristic measures of fractional point processes. We analyse spontaneous locomotor activity of healthy individuals recorded during a week of regular sleep and a week of chronic partial sleep deprivation. Behavioural symptoms of lack of sleep can be evaluated by analysing statistics of duration times during active and resting states, and alteration of behavioural organization can be assessed by analysis of power laws detected in the event count distribution, distribution of waiting times between consecutive movements and detrended fluctuation analysis of recorded time series. We claim that among different measures characterizing complexity of the actigraphy recordings and their variations implied by chronic sleep distress, the exponents characterizing slopes of survival functions in resting states are the most effective biomarkers distinguishing between healthy and sleep-deprived groups.
\end{abstract}

\maketitle
{\bf SUBJECT AREA}: mathematical physics, theory of complex systems, biomathematics\\
\section{Introduction}
Despite numerous studies indicating anomalous temporal statistics and scaling in spontaneous human activity and interhuman communication \cite{Barabasi,Bickel,Anteneodo,Marin,Haimovici,Dante}, there is much on-going discussion on the origin and universality of observed statistical laws. Behavioural processes are frequently conveniently characterized in terms of stimulus-response approach \cite{Marin,Proekt}, by adapting the same systematically repeated external sensory protocol, which allows to estimate the statistics of subject's responses.
 In a more general approach, in which brains are conceived as information processing input-output systems \cite{Enzo,Palva}, the observed self-similar temporal patterns of non-stimulated spontaneous neuronal activity can be determined by analysing spatiotemporal statistics of location and timing of neural signals. Similar to scale-free fluctuations detected in psychophysical time series, also dynamics of collective neuronal activity at various levels of nervous systems exhibit power-law scalings. Remarkable scale-free fluctuations and long-range correlations have been detected on long time scales (minutes and hours) in data recorded with magneto- and electroencephalography \cite{Palva}, and have been attributed to the underlying dynamic architecture of spontaneous brain activity discovered with functional MRI (fMRI) and defined by correlated slow fluctuations in blood oxygenation level-dependent (BOLD) signals.
 
 On the other hand, negative deflections in local field potentials recorded at much shorter time scales (milliseconds) have been shown to form spatiotemporal cascades (neuronal avalanches) of activity, whose size (amplitude) and lifetime distributions are again well described by power laws. These power-law scaling behaviours and fractal properties of neuronal long-range temporal correlations and avalanches strongly suggest that the brain operates near a critical, self-organized state \cite{Enzo} with neuronal interactions shaping both, temporal correlation spectra and distribution of signal intensities. It seems thus plausible to further investigate timing, location and amplitudes of such cascades to gain information about underlying patterns of brain rhythms and to identify characteristics of stochastic spatial point processes which can serve as reliable models of the governing dynamics. Also, if the behaviour can be described and interpreted as resulting interface between brain dynamics and the environment, scale invariant features may be expected to emerge in human cognition \cite{Kello}  and human motion \cite{Dante,Nakamura,NakamuraPLOS,Stang,Teicher}.
 
Some neurological and psychopathic diseases such as Parkinson's disease, vascular dementia, Alzheimer's disease, schizophrenia, chronic pain and even sleep disorders and depression are related to abnormal activity symptoms \cite{Pan,Pankwak,Kim}. So far, there are many non-unique evaluative measures used in clinical practice to determine severity of these disorders or the effect of applied drugs. The challenge thus remains to what extent correlations during resting state (spontaneous) activity are altered in disease states and whether a set of characteristic parameters can be classified unambiguously to describe statistics of healthy versus unhealthy mind states and spatiotemporal organization of such disrupted brain dynamics. 


An objective measure of differences in sleep durations and effects of sleep deficit on restlessness and impulsivity in human activity is based on {\it actigraph recording}. The instrument is used in sleep assessment to discriminate between stages of sleep and wake through documented body movements \cite{Teicher}. The analysis of actigraphy data is frequently considered a cost-effective method of first choice, used both in clinical research and practice.

The aim of this study is to apply various complexity measures for large activity datasets and to quantify use of advanced statistical methods for activity data. Our analysis is based on assumption that a typical actigraph time series that monitor records of movements per fixed time period of known duration can be analysed as a stochastic point process. We establish a time-inhomogeneity of such a renewal process and describe its fractional character by analysing counting number distributions. The renewal process is specified by probability laws describing waiting times or inter-arrival (inter-event) times \cite{Cox, Mainardi,Gardiner}. The survival probability $C(a)\equiv Prob(T\geq a)=\int_{a}^{\infty}P_T(\tau)d\tau$ is a common measure that relates the (mean) time of discharge (or escape time) to the probability density function $P_T(t)$. In contrast to the Poisson case, for which a characteristic mean time of escape exists, fractional point processes may possess $C(a)$ and $P_T(\tau)$ functions which do not decay exponentially but algebraically. As a consequence of the power-law asymptotics, the process is then non-Markovian with a long memory, which may result in a lack of characteristic time-scales for relaxation.

The paper is organized as follows: In subsequent sections the method of data collection is presented (Section II), followed by a brief description of a stochastic point process (Section III) which serves as a model for further statistical analysis. Section IV demonstrates our results obtained by use of different complexity measures. The findings are concluded in Section V. 

\section{Experiment: Actigraphy recordings}
\label{sec:acti}
An actigraph is a watch-like device worn on the wrist that uses an accelerometer to measure any slight body movement over the investigated period of time. This instrument applies simple algorithms (time-above-threshold, zero-crossing, and digital integration) to summarize the overall intensity of the measured movement within consecutive time-periods (usually 1-2 minutes long), so that the events recorded represent activity counts. The most general approach to analyse such a recording is to reduce the time series of measurements to a summary statistic such as sleep/wake ratios, sleep time, wake after sleep onset, and ratios of night-time activity to daytime or total activity \cite{Teicher,Pan,Pankwak}. While such summary measures allow for hypothesis testing using classic statistical methods \cite{Teicher,Oginska,Sun,Matuzaki}, large amounts of relevant information  are disclosed in those complex data when more advanced tools of quantification (cf. Section \ref{sec:model}), like fractal analysis, are applied \cite{Holloway,Stang,Pan,Ochab,Nakamura,Kim}. In particular, actigraphy studies by Sun et al. \cite{Pan} indicated that the scaling exponent of the power law detected in temporal autocorrelation of activity significantly correlates with the severity of Parkinson's disease symptoms. Similarly, universal scaling laws have been found in locomotor activity periods of humans suffering from major depressive disorders \cite{NakamuraPLOS} and individuals subject to sleep debt \cite{Ochab,Holloway}. The disruption of the characteristic universality classes of such laws has been further addressed by Proekt et al. \cite{Proekt} in studies on dynamics of rest and activity fluctuations in light and dark phases of the circadian cycle.

\begin{figure}[h!]
\includegraphics[width=0.98\columnwidth]{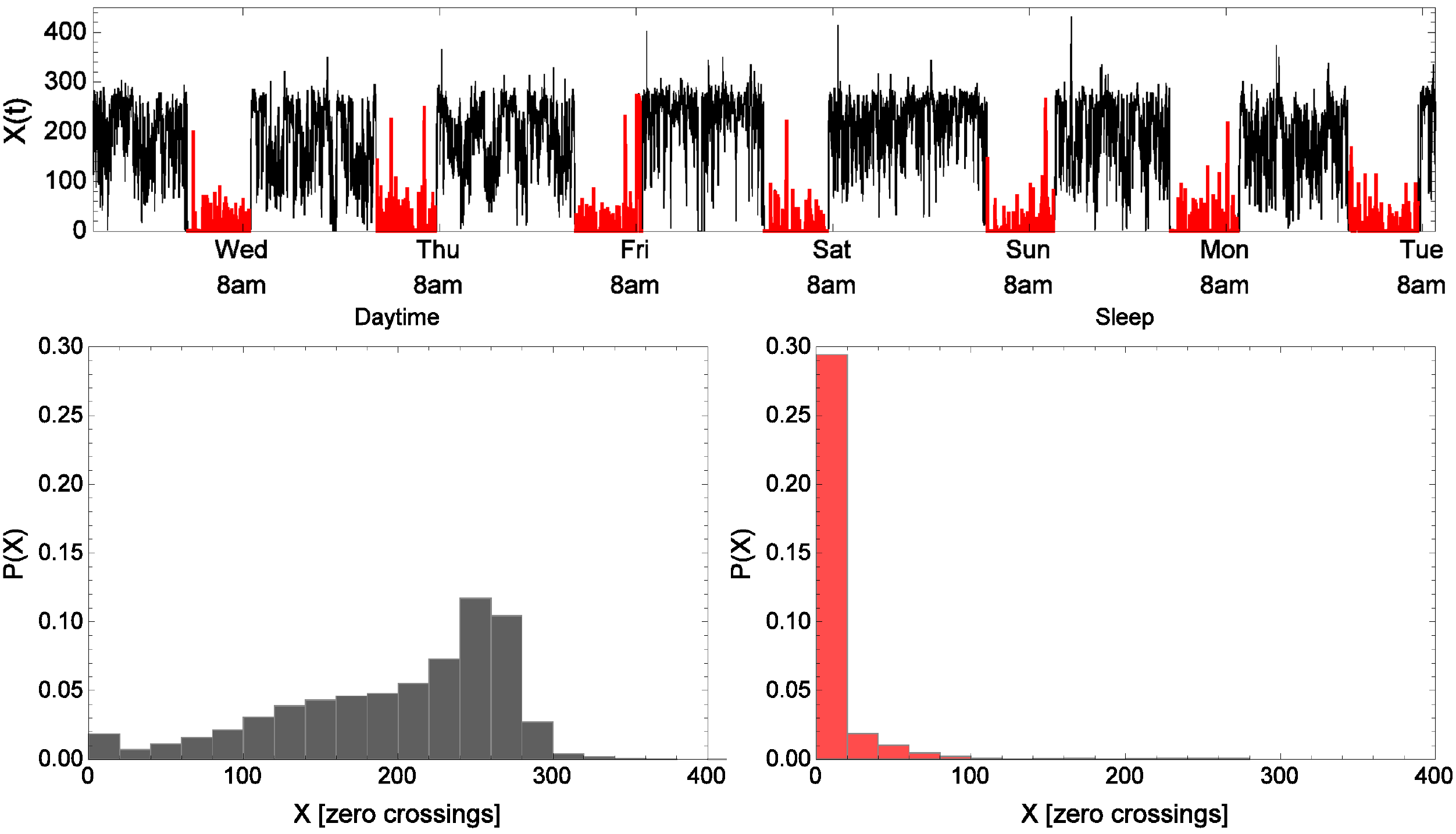}
\caption{\label{fig:rawdata}Time series for one-week recording of actigraphy in the ZCM mode for a typical subject in RW condition. Upper panel: colours in the graph reflect daytime (black) and night (red) recordings following the circadian rhythm. Lower panels depict normalized frequency histograms of the number of movements in a minute derived from the raw actigraphy recordings. Unlike histograms of activities at night, which contain a considerable amount of zero-recordings (no movement) within 1 min intervals, histograms of awake activity reflect higher dispersion and multi-modality.}
\end{figure}

\begin{figure}[h!] 
\includegraphics[width=0.6\columnwidth]{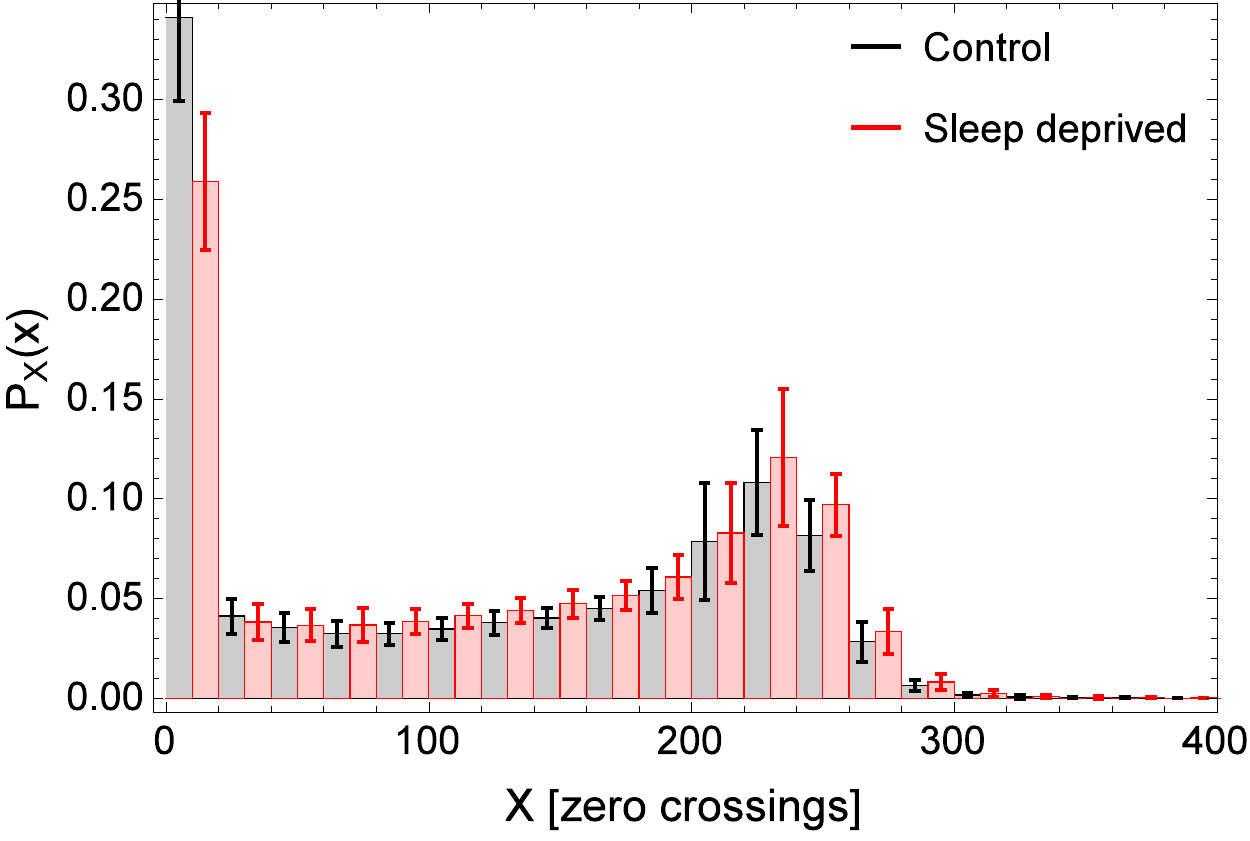}
\caption{\label{fig:activity} Raw actigraphy recordings: Figure displays a normalized histogram of activity in the ZCM mode averaged over all subjects (error bars are standard deviations). Horizontal axis corresponds to the number of activity counts within one minute intervals, while vertical axis to normalized frequencies of counts. Black bars reflect frequency values for the control group (individuals in rested wakefulness; RW), while red bars the frequencies for subjects undergoing sleep deprivation (SD). The activity counts show a clear bimodal distribution with peaks characterizing maxima of low- and high-activity subdistributions (cf. Fig.~\ref{fig:rawdata}).}
\end{figure}

Here, the actigraphy measurements were performed on 18 healthy individuals over one week of their normal life with either full recovery sleep according to individual needs, i.e., at rested wakefulness (RW or control group), or with a daily partial sleep deprivation (SD), with a two-week gap in between the two conditions. 
During the sleep deficit week, the participants were asked to shorten their sleep by 33\% (by 2 hours 45 minutes on average) of their "ideal sleep" by delaying bed-time and using an alarm clock in the morning.
The precise length of the restricted sleep was calculated individually for each participant.
Half of the subjects began with the RW phase followed by the SD phase, while the other half had the order reversed. 
Movement tracking was recorded with Micro Motionlogger SleepWatch (Ambulatory Monitoring, Inc., Ardsley, NY), worn on the participant's non-dominant wrist. The data were collected in 1-min epochs in the Zero-Crossing Method (ZCM) mode,
which counts the number of times per epoch that the activity signal level crosses a zero threshold \cite{Ochab}. 
An example of the raw time series, as well as resulting activity histograms are displayed in Fig. \ref{fig:rawdata}. The histograms for RW and SD groups are compared in Fig. \ref{fig:activity}. Both, night-time and daytime measurements were analysed and the data collected are provided in a supplementary material (see website of the Department of the Theory of Complex Systems, Jagiellonian University, http://cs.if.uj.edu.pl/jeremi/Actigraphy2013/).

The circadian cycle of both groups differs substantially: while RW individuals have relatively long ``nights" and short ``days", members of the SD group are characterized by a reversed motif of longer ``days" and shorter ``nights", which clearly influences their activity/rest patterns.
To overcome this problem we normalized the ``days" and ``nights" of both groups to the same length. The resulting time series have been statistically analysed as described previously in \cite{Ochab}. In addition, in this study, complexity analysis has been extended to describe the statistics of events' counts as a function of the observation time.

\section{Model: Spiking patterns and measures of deviation from uniformity}
\label{sec:model}
 
 Many pulse signals observed in nature can be described in terms of stochastic point processes, i.e., sequences of random points $\{t_i\}$ arriving in time \cite{Bialek}. The simplest realization of such signals can be a ``spike train" composed of delta peaks arriving at random times $Y(t)=\sum_{t_i}\delta(t-t_i)$. The average value of such a signal is given by a spike rate $r(t)$, or the firing probability density (firing probability per unit of time):
 \begin{equation}
 \lim_{\Delta t\rightarrow 0}\frac{1}{M\Delta t}\sum^M_{i=1}\int ^{t+\Delta t/2}_{t-\Delta t/2}dt'Y_i(t')\equiv\left\langle Y(t)\right\rangle = r(t).
 \end{equation}
 Here $Y_i(t)$ stands for a given realization of the firing process. The firing rate can be thus obtained by observing at least one spike in the time interval $[t-\Delta t/2, t+\Delta t/2]$, dividing this probability by $\Delta t$ and taking the ensemble average (over $M$ realizations).
Accordingly, correlation function of the spike train is defined as
\begin{eqnarray}
\left\langle Y(t_1)Y(t_2)\right\rangle\equiv C(t_1,t_2)=\nonumber \\
= \lim_{\Delta t\rightarrow 0}\frac{1}{M\Delta t}\sum^M_{i=1}\int ^{t_1+\Delta t/2}_{t_1-\Delta t/2}dt_1'\int ^{t_2+\Delta t/2}_{t_2-\Delta t/2}dt'_2Y_i(t_1')Y_i(t'_2)
\end{eqnarray}
and for a stationary process it depends on the time difference $t_1-t_2$ only \cite{Gardiner}.
Vanishing width of impulses allows one to construct another descriptor of the spike train, i.e., a counting process $N(t)$ which gives the count of impulses observed in a given time window 
\begin{equation}
N(t)=\int^t_0 dt' Y(t').
\end{equation}
If the spike sequence depends only on the homogeneous rate 
\begin{eqnarray}
r=\lim_{\Delta t\rightarrow 0}\frac{1}{\Delta t}{Prob\{\Delta N(t,t+\Delta t)=1\}},
\end{eqnarray}
and the events appear independently, the resulting distribution of counts can be described by the probability $P(N, t)=p_n(t)\delta (N-n)$ satisfying a balance equation $\dot{p}_n=rp_{n-1}-rp_n$. When  $p_n(0)=0$ for $n\ge 1$ and $p_0(0)=1$, the recurrence leads to the general solution $p_n(t)=\frac{(rt)^n}{n!}e^{-rt}$. In this scenario, the probability density function (PDF) of inter-event times has the exponential form $P_{T}(\tau)=re^{-r\tau}$ which is a signature of spikes following each other on average at $\left\langle\tau\right\rangle=1/r$ intervals. Since dispersion of $\tau$ is finite in this case, very long inter-event times are exponentially rare.

Passing the $\delta$-pulse signal of intensity $\sigma$ through a linear filter 
\begin{eqnarray}
\tau \frac{dX}{dt}=-X+\sigma Y(t)
\end{eqnarray}
results in ``shot noise" with exponentially decaying pulses
\begin{eqnarray}
X(t)=\frac{\sigma}{\tau}\sum_i\Theta(t-t_i)\exp[-\frac{t-t_i}{\tau}].
\end{eqnarray}
The above equations represent a basic model \cite{Brunel,Rudolph} for synaptic noise with $X(t)$ standing for a neuron's membrane potential and $\tau$ denoting relaxation time of the activated neuron. In more general terms one can consider a stochastic process $X(t)=\sum_jc_jx_j(\omega_j (t-t_j))$ resulting from superposition of many statistically identical stochastic signals $x_j$ transmitted from independent sources whose intensity $c_j$, frequencies $\omega_j$ and  "initiation" times $t_j$ may be also considered random variables \cite{Klafter}. Such superposition models are salient in systems carrying information traffic and well qualify long-range dependence, power-law scaling, $1/f$ noise and anomalous relaxation \cite{Eliazar}. Since extensive research shows that also brain dynamics\textemdash from cultured single neurons to whole brains\textemdash exhibit spatiotemporal self-similarity \cite{Werner}, it is tempting to assume a direct relation between  the scaling laws of behavioural (macroscopic) and neuronal (microscopic) fluctuations \cite{Werner,Palva} and to approach them theoretically by similar models.

Poissonian randomization of the parameter set $\{(c_j,\omega_j,t_j)\}$ on the half space
 $(-\infty, \infty)\times (-\infty,\infty)\times (0,\infty)$ establishes a mapping of the microscopic signal to the output "macroscopic" process $X(t)$. As discussed in \cite{Klafter,Eliazar}, if the Poissonian intensity follows algebraic Pareto laws, the aggregate "output" signal can exhibit amplitude-universal and temporally universal statistics independent of the pattern of generic signals $x_i(t)$. 
 The emerging output patterns may then display  inter-event time PDFs characterized\textemdash unlike their Poisson (regular) counterparts\textemdash by power-law tails $P_{T}(\tau)\propto \tau^{-\alpha}$ . This in turn implies ``burstiness" with short time intervals of intense spiking separated by long quiescent intervals, contrasting strongly with a uniform spiking pattern of a Poisson processes.

A general type of clustering processes stemming from the Poisson class has been introduced by Neyman \cite{Neyman,Vlad,Bickel}. In short, the Neyman-Scott process is a mixture of a Poisson (parental) process of cluster centres with a characteristic intensity $r(t)$ with another, statistically independent (daughter) stochastic process which governs distribution of points within the cluster. 
The Laplace transform (moment generating function) of a random variable representing a compound signal $X_{N(t)}\equiv\sum_i^{N(t)}X(i)$, conditioned on Poisson distribution of random variable $N$ takes then the form of $\Phi_{X_{N(t)}}(s)\equiv \left\langle X_{N(t)}^s\right\rangle=
\Phi_{N(t)}(\Phi_{X_i}(s))$. It is easy to check that dispersion (or so called Fano factor), defined as
\begin{equation} 
FF\equiv \frac{\left\langle X^2_{N(t)}-\left\langle X_{N(t)}\right\rangle^2 \right\rangle }{\left\langle X_{N(t)}\right\rangle }
\label{Fano}
\end{equation}
for such a distribution is greater than 1, the value characteristic of a simple Poisson distribution with uncorrelated events. Accordingly, dispersion of experimental distributions, derived by use of Eq.(\ref{Fano}), serves as an indicating measure for clustering.
Here, the point process is defined as an activity crossing a predefined threshold (see the description of zero-crossing mode in Sec. \ref{sec:acti}). This threshold is preset by the producer of the actigraphs; we do not have access to it. It should not be mistaken for the threshold used to define activity and rest periods in Sec. \ref{sec:survival}.
Note that since the events are recorded within 1-minute bins, we do not have access to the time variable below the bin length.

In order to detect variability of the difference in the number of events
occurring in successive intervals $t$, another measure\textemdash the Allan variance\textemdash is frequently used
\begin{equation}
AF\equiv\frac{\left\langle [X_{N(t)+1}-\left\langle X_{N(t)}\right\rangle]^2\right\rangle }{\left\langle 2X_{N(t)}\right\rangle }
\label{Allan}
\end{equation}

In case of the fractal-rate stochastic point processes, for which the times between adjacent events are random variables drawn from fractal distributions, one observes a hierarchy of clusters of different durations \cite{Vlad,Bickel,Kaulakys,Laurson} and both Fano and Allan factors deviate from unity.

In what follows, we consider measures suitable to discriminate basic properties of the point process representing events' count in actigraphy recording. First, we analyse frequency histograms $P_X(x)$ of signal intensity and statistics of inter-event intervals. To further detect character of 
the event count process, we evaluate variations of rate $r(t)$ in expanding time windows checking for event clustering. Analysis of variance of counts, based on derivation of Fano and Allan factors allows to detect non-Poissonian character of the counting process and identify presence of correlations in recorded sequences.

\section{Results of statistical analysis}
\subsection{Frequency histograms for distributions of activity increments}

\begin{figure}[htb]
\includegraphics[width=0.9\columnwidth]{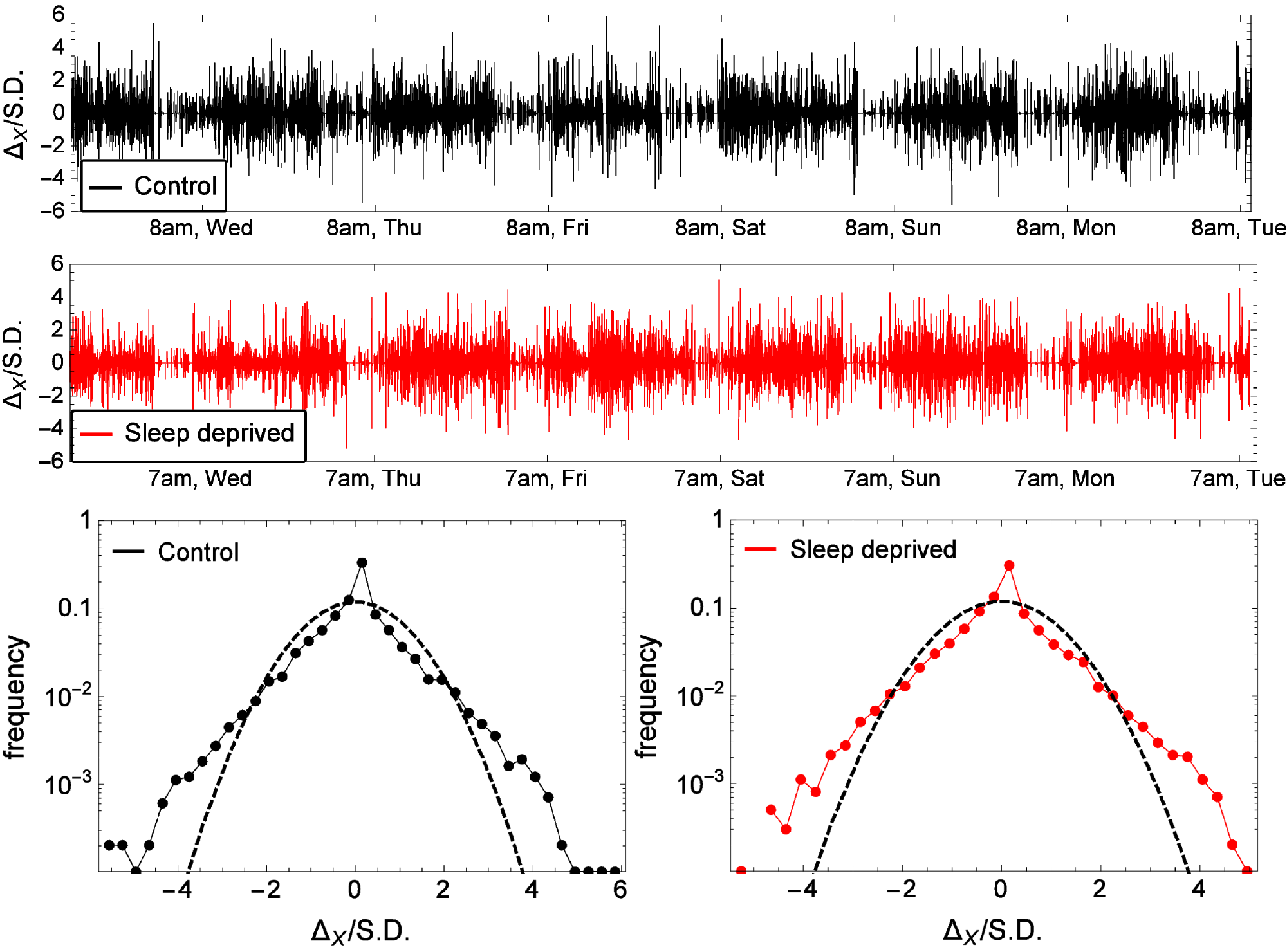}
\caption{Time series of activity increments $\Delta_X$ for a typical subject. Lower panel depicts density distribution $P_{\Delta_X}(\frac{\Delta_X}{S.D.})$ for RW and SD condition. The dashed line is a Gaussian density with the same variance.}
\label{rate}
\end{figure}

In order to detect dynamics of fluctuations in the intensity $X(t)$, (cf. Fig.\ref{fig:rawdata}), we have analysed increments $\Delta_X = X(t+1)-X(t)$ for RW and SD subjects separately, and computed density distributions $P_{\Delta_X}$ presented in Fig.~\ref{rate}.
The plots have been standardised and compared to a Gaussian density with the same mean and variance.
There is a clear discrepancy between the Gaussian statistics and the observed densities of activity fluctuations. They display much heavier tails in intensity distribution than their Gaussian counterparts, indicating that large fluctuations in intensity of the signal are more frequent than predicted from a standard (classical) central limit theorem.

\subsection{Timing of events: Rate analysis}
\label{sec:FF}
\begin{figure}[htb]
 \includegraphics[width=0.48\columnwidth]{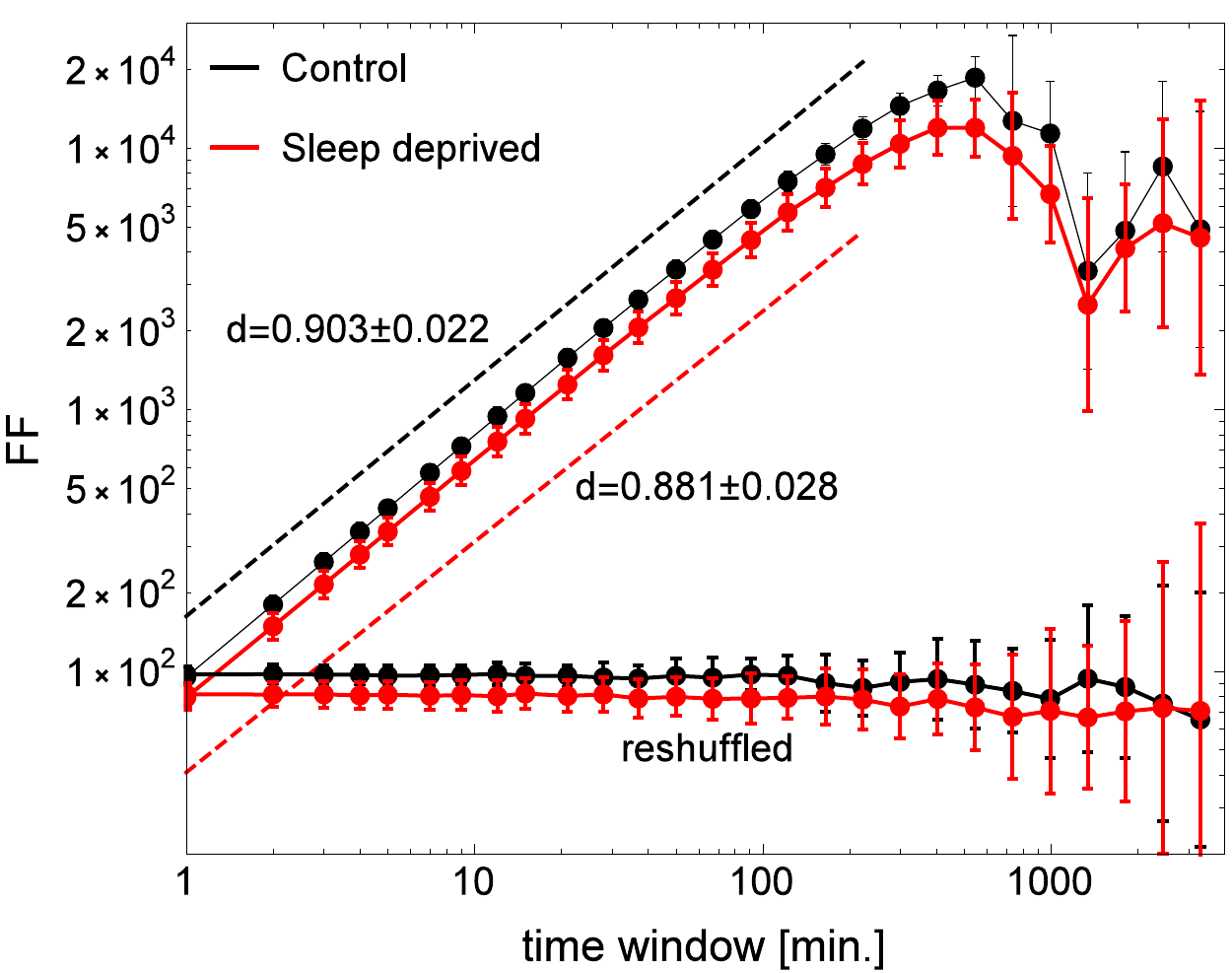}
 \hfill
 \includegraphics[width=.49\columnwidth]{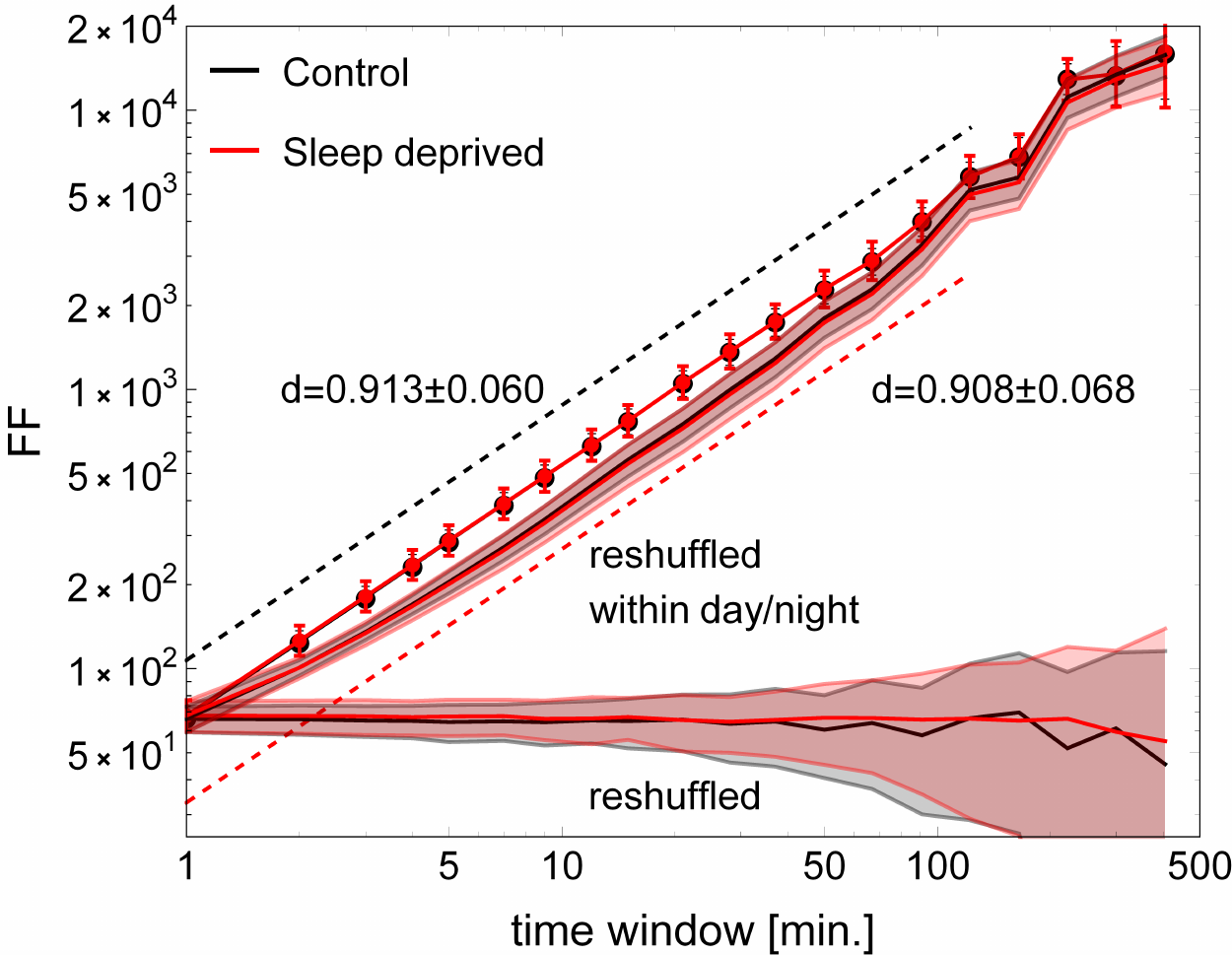}
\caption{\label{fig:fano24} Fano factor (FF) of the 7-day raw time series for the control (black) and sleep deprived (red) group. Each point is an average FF over 18 subjects; the error bars and shaded regions indicate standard deviations. The range and result of linear fit is indicated by the dashed lines. The lower horizontal curves show FF for time-reshuffled data.
Left: untrimmed 24h days; the slopes of linear fits of the two curves are not significantly different, but the $y$-intercepts are.
Right: days trimmed to 4h sleep followed by 14h daytime activity; there is no significant difference between the intercepts, which implies that the difference visible for 24h-days can be accounted for solely by the different lengths of day/night activity.
Reshuffling the data within day/night compartments, shows that the clustering of activity indicated by FF can be introduced by day/night cycle. We scrutinize this effect in Fig. \ref{fig:fanoDN}.}
\end{figure}

\begin{figure}[htb]
 \includegraphics[width=0.49\columnwidth]{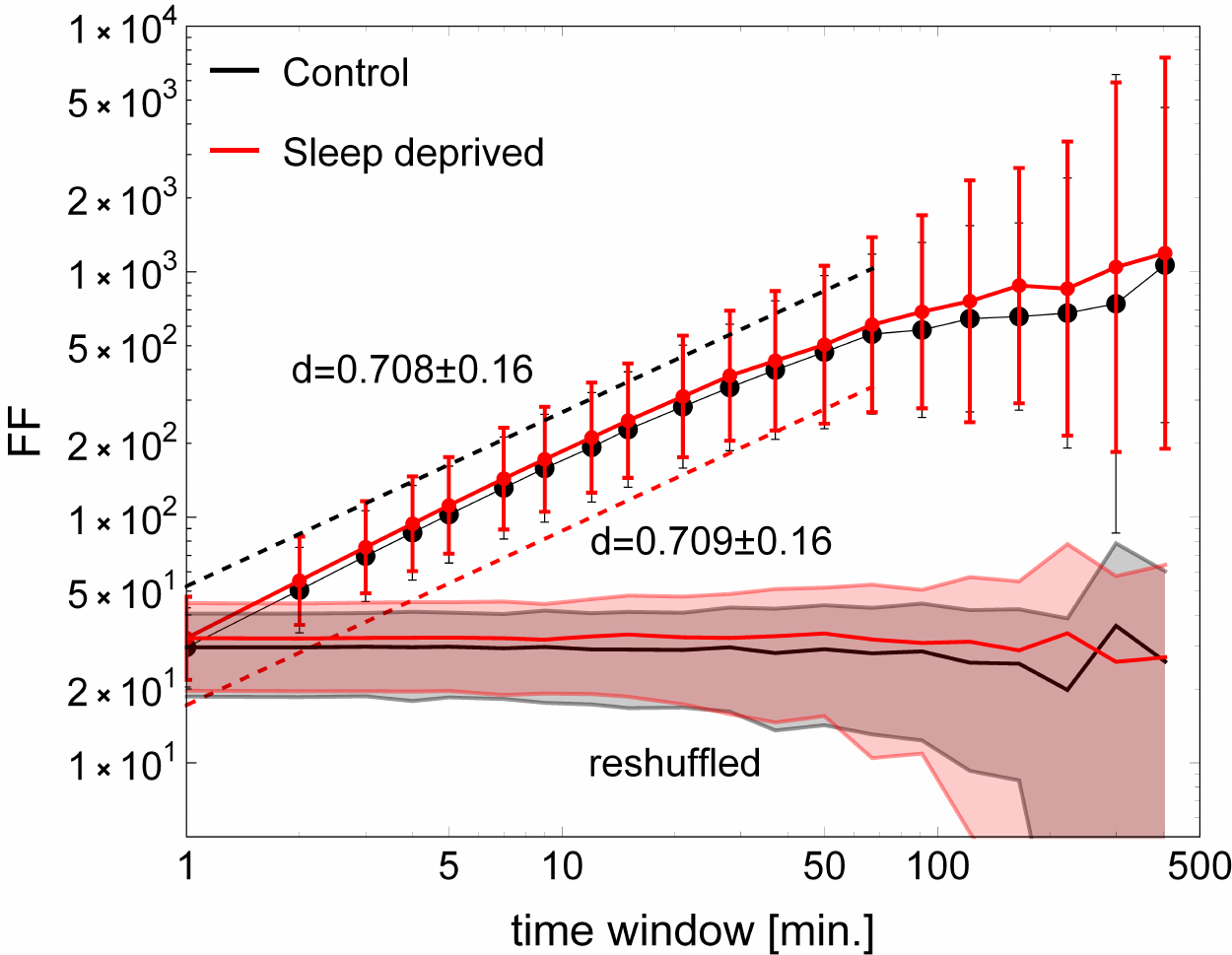}
 \hfill
 \includegraphics[width=.49\columnwidth]{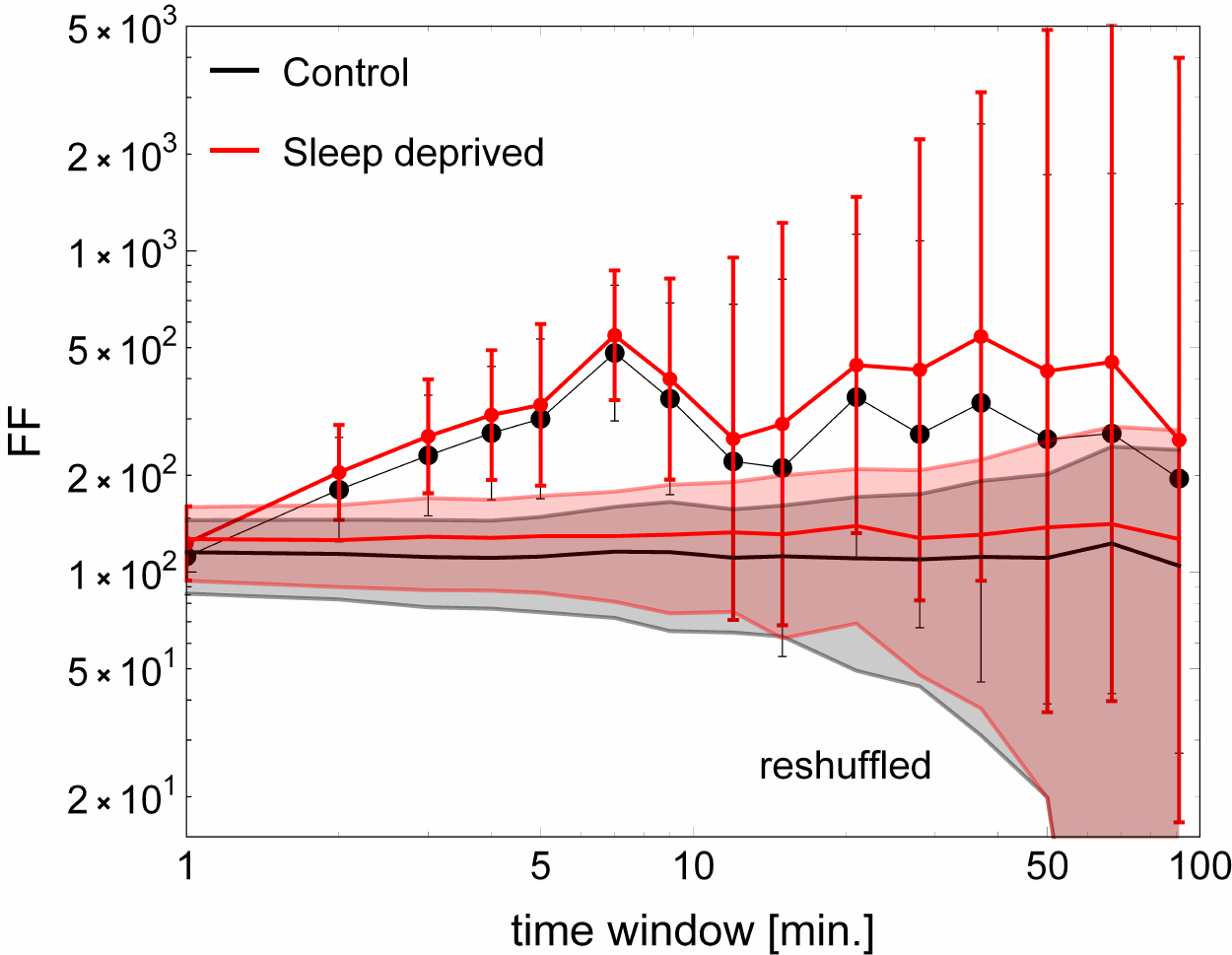}
\caption{\label{fig:fanoDN} Fano factor as in Fig. \ref{fig:fano24} controlled for time of day. Left: FF for 14h daytime; the power-law behaviour remains, although with a markedly smaller exponent. Right: FF for 4h night-time; significantly non-random clustering of activity can be observed only for very short windows.
There is no significant difference between control and SD curves.}
\end{figure}

\begin{figure}[h!]
 \includegraphics[width=0.49\columnwidth]{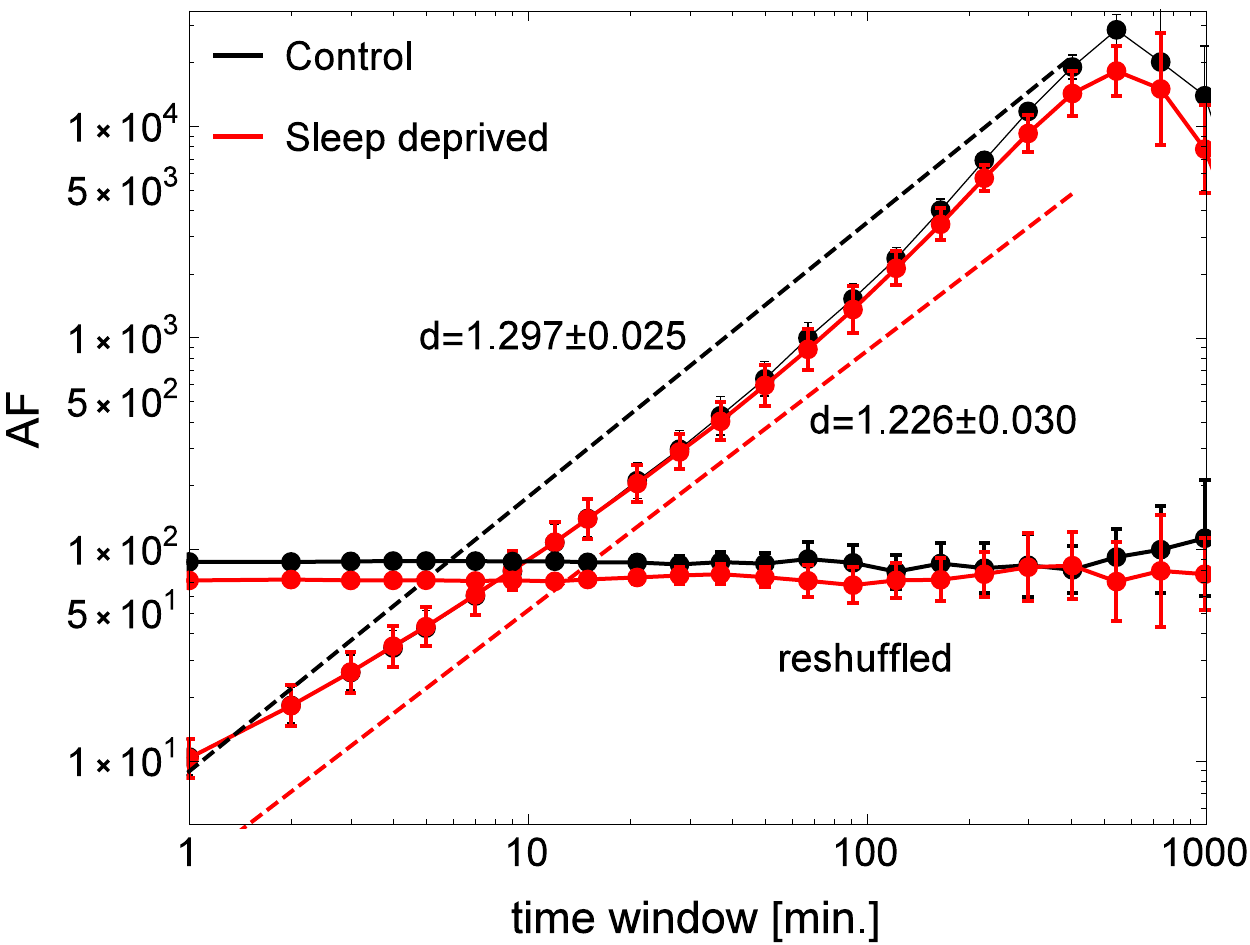}
 \hfill
 \includegraphics[width=0.48\columnwidth]{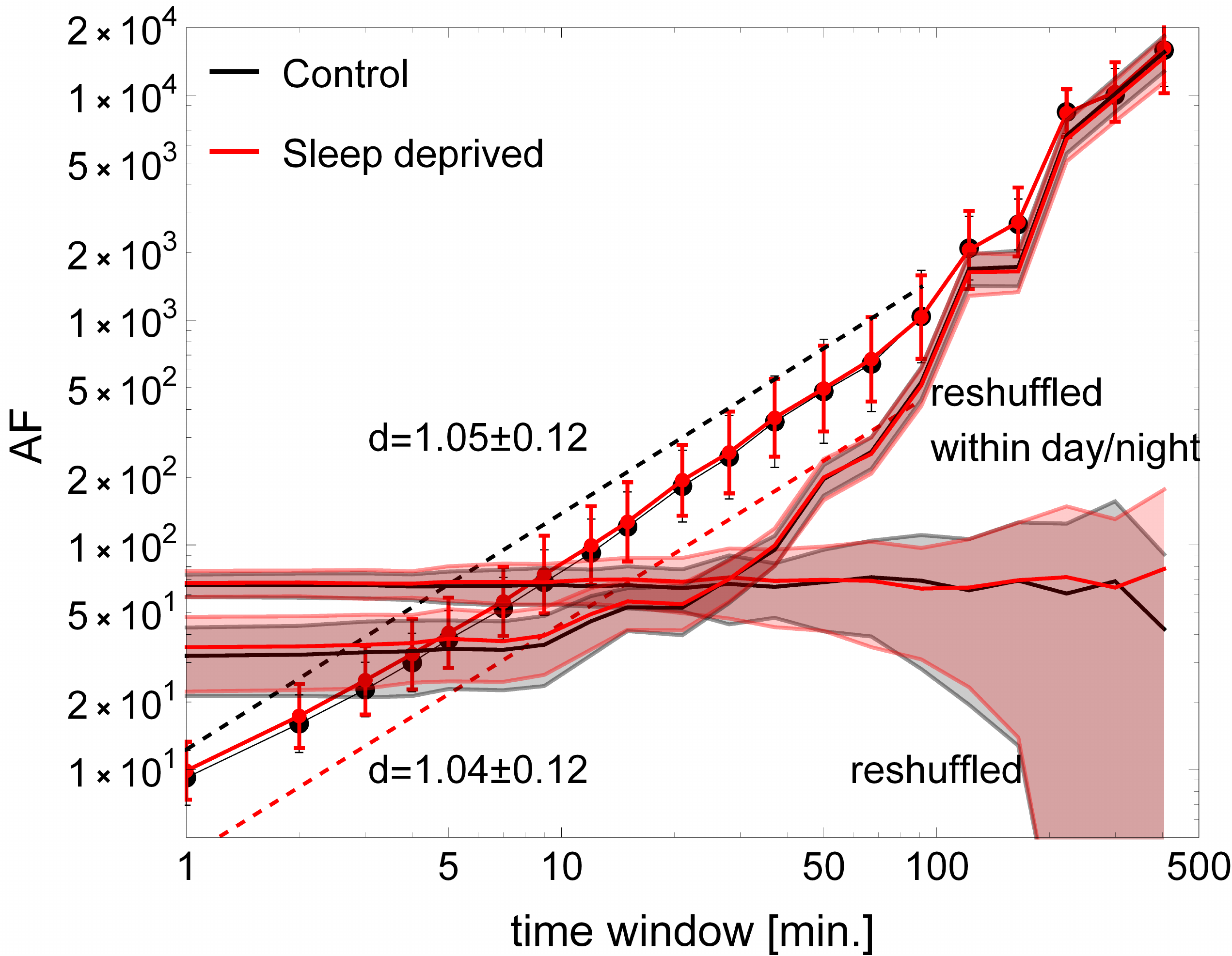}
\caption{\label{fig:allan24}
Allan factor (AF) of the 7-day raw time series for the control (black) and sleep deprived (red) group, as in Fig. \ref{fig:fano24}. Left: untrimmed 24h days; the $y$-intercepts of linear fits of the two curves are not significantly different, but the slopes are. Right: days trimmed to 4h sleep followed by 14h daytime activity. 
There is no significant difference between control and SD curves, which implies that the differences for 24h-days can be accounted for solely by the different lengths of day/night activity, as previously observed for the FF measure.
Reshuffling the data within day/night compartments, shows that the day/night cycle introduces clustering of activity between windows longer than $10$ min. We scrutinize this effect in Fig. \ref{fig:allanDN}.}
\end{figure}

\begin{figure}[h!]
 \includegraphics[width=0.49\columnwidth]{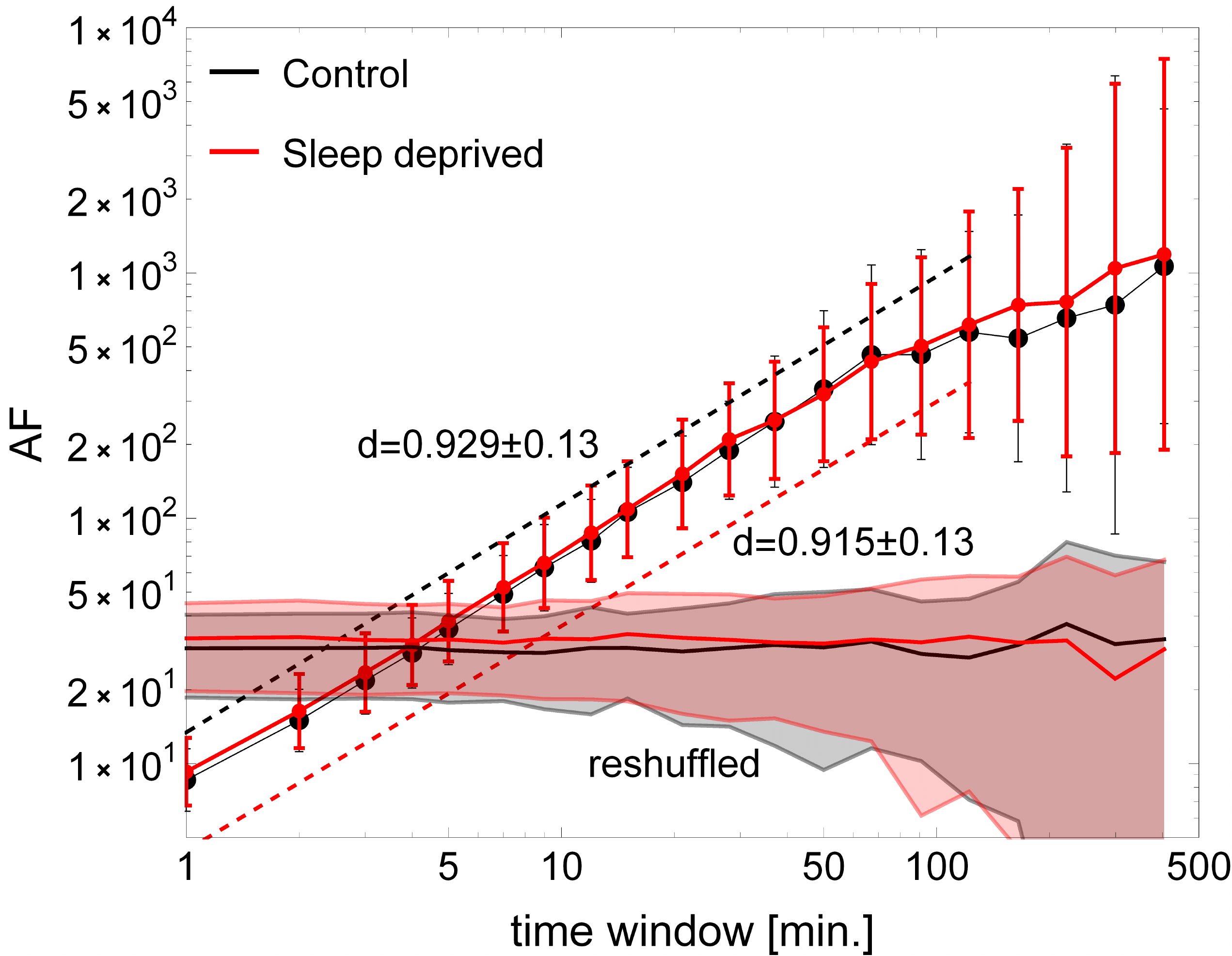}
 \hfill
 \includegraphics[width=0.49\columnwidth]{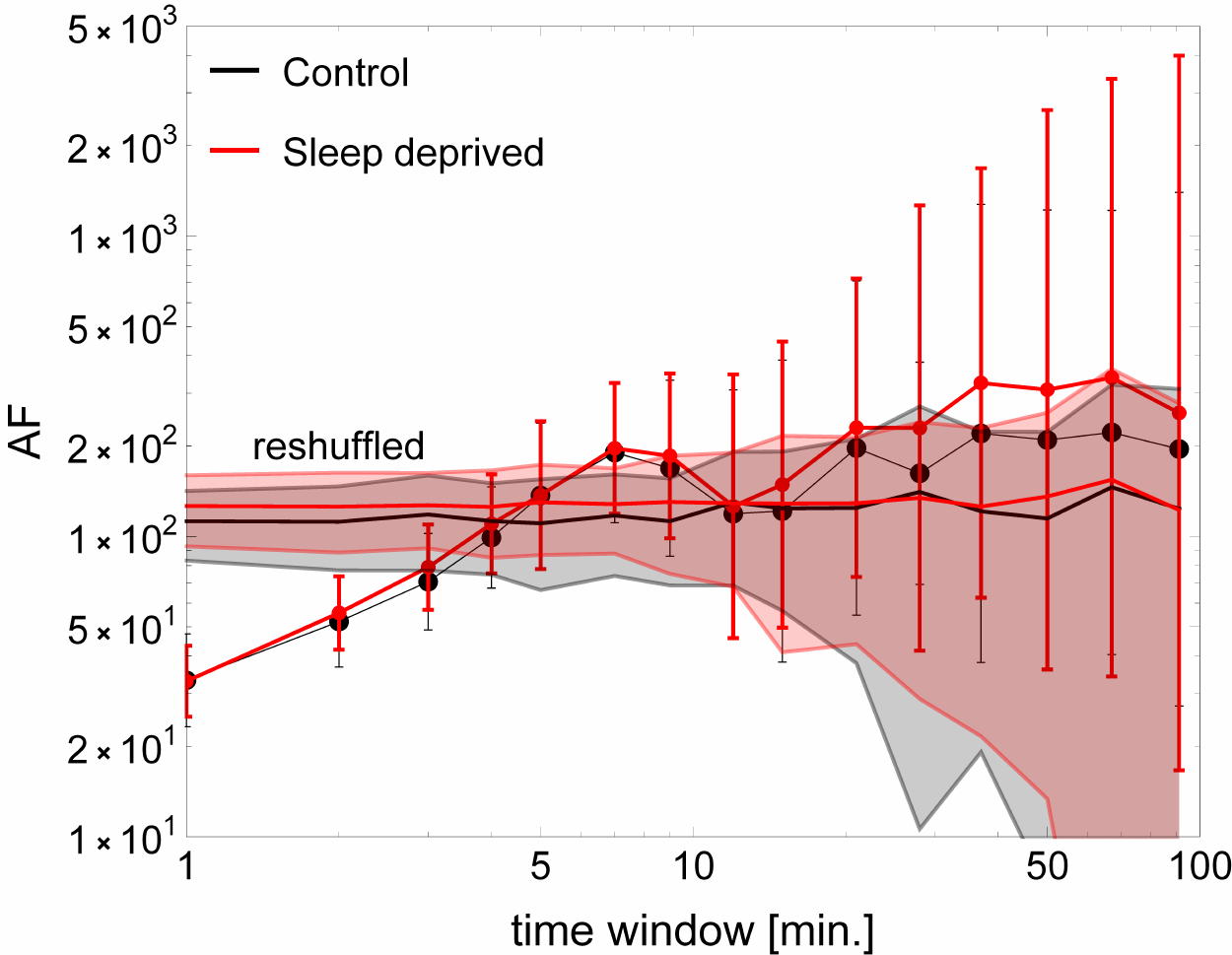}
\caption{\label{fig:allanDN}
Allan factor as in Fig. \ref{fig:allan24} controlled for time of day. Left: AF for 14h daytime; the power-law behaviour remains, although with a slightly smaller exponent. Right: AF for 4h night-time; clustering of activity can be observed only between very short successive windows.
There is no significant difference between control and SD curves.}
\end{figure}

The degree of non-homogeneity of the event counting process $N(t)$ has been approached by considering cumulative number of events as a function of time.
The fluctuations in the number of events $N(t)$ crossing a predefined activity threshold $X_{th}$ has been analysed utilising Fano (FF, Eq.~\ref{Fano}) and Allan (AF, Eq.~\ref{Allan}) factors depicted in Figs. \ref{fig:fano24}-\ref{fig:allanDN}. The averages in the definitions of FF and AF have been calculated by taking consecutive, non-overlapping windows and disregarding the last one, if its length was smaller than the counting time $T$.
As can be inferred from the displayed data sets, the quantities: FF measuring the event-number variance divided by event-number mean and AF reflecting discrepancy of counts in consecutive time windows, both strongly deviate from unity during daytime (left panels in Fig. \ref{fig:fanoDN} and \ref{fig:allanDN}), thus indicating a super-Poisson statistics $N(t)$ with clustering of events in the observation time. Since the obtained fractal exponents are close to one, which is the maximal exponent value that FF can detect, AF seems better fitted for the analysis, as it can detect exponent values up to three.

Both factors increase as $T^d$ within counting time $T\in[1, 500]$. Reshuffling time stamps in the series breaks down inter-event correlations and results in reduced and uniformly distributed values of FF and AF, as expected \cite{Anteneodo,Bickel}. Separate analysis performed for daytime and night-time data does not discriminate between FF and AF values derived for RW and SD groups. Also, no significant difference in estimated $d$ exponents for those two groups is observed when the analysed time series are all standardized to the length covering 4h of sleep followed by 14h of daytime activity (alternatively, 14h of daytime activity followed by 4h of sleep).

The FF curves for raw time series of the whole week of recording (Fig.~\ref{fig:fano24}) were fitted with least-squares linear regression on logarithmic data in the range indicated by the dashed lines (17 data points, with window sizes: 1-222; the fitting range  was determined by dropping the rightmost points until the adjusted $R^2$ coefficient stabilised).
A leave-one-out cross-validation yields sample standard deviation of the fitted slope: $0.024$ ($0.030$) for RW (SD), which is less than $10\%$ larger than the standard error of the fit of the subject-averaged curve (Fig.~\ref{fig:fano24}, left panel). This ensures that the fitting regions were chosen reliably.
The fitted lines are not significantly different in terms of slopes (the values of $d$ together with the standard error of the fit are reported in the figure; the difference is below 1 SE), but after subtracting the common slope they are different in terms of the $y$-intercept ($98.2\pm 1.1$ and $82.5\pm 1.1$ for RW and SD, respectively; a two-sided $t$-test for the difference between the intercepts yields $p=0.015$).
The same vertical shift is present between time-reshuffled data. Hence, the difference between the intercepts can be accounted for solely by the different lengths of day/night activity. To avoid that, the data can be trimmed to the same day/night lengths, which reduces the shift as in Fig.~\ref{fig:fano24} (right).

To assess the inter-subject variability, the standard deviation of slopes fitted for individual subjects (averaged over days) was calculated yielding $\leq 0.030$ for both untrimmed and trimmed time-series. The intra-subject variability, can be described by standard deviation of the fitted slopes for a given subject but for different days; it yields $\leq 0.062$. This means that the measurements across subjects and across days in the experiment can be expected to vary with similar magnitude; as a consequence, the subject-averaged curves should be indicative of the whole sample and, for clarity, we do not show the individual subject values.

As a cautionary note, let us remark that the slopes fitted for individual subjects are different between the RW and SD groups under a paired $t$-test, but this difference disappears after the trimming as well. A paired $t$-test in such cases, however, may produce a false positive conclusion, because it does not take into account error of the fit of the slopes. Slope of an averaged curve  thus provides more conservative test conclusions.

The lines fitted for AF curves (Fig.~\ref{fig:allan24}; 20 data points, with the window sizes ranging 1-404) are not different in terms of their intercepts ($5.4\pm 1.1$ for RW and $6.1\pm 1.1$ for SD, where the numbers correspond to the fitted intercept and standard error of the fit; the difference between RW and SD is around 1 SE), but after subtracting the common intercept they are different in terms of the slope (see values in the figure; a two-sided $t$-test for the difference between the adjusted slopes yields $p=0.035$). The lower horizontal curves in Fig.~\ref{fig:allan24} show AF after randomly reshuffling time stamps of the raw data, resulting in a more pronounced difference between the curves. Again, those differences can be accounted for solely by different time-lengths of day/night activity, so that data trimming is necessary.

Observation of AF curves starting to separate only for large window sizes, with the FF and reshuffled AF curves separated for all window sizes, can be explained by the fact that AF measure is based on increments between neighbouring windows. This implies that AF curves would separate only when the windows are big enough to often contain the ``edge" between a day and a night.

\subsection{Detrended fluctuation analysis and Hurst exponent}

\begin{figure}[h]
\includegraphics[angle=0,scale=0.38]{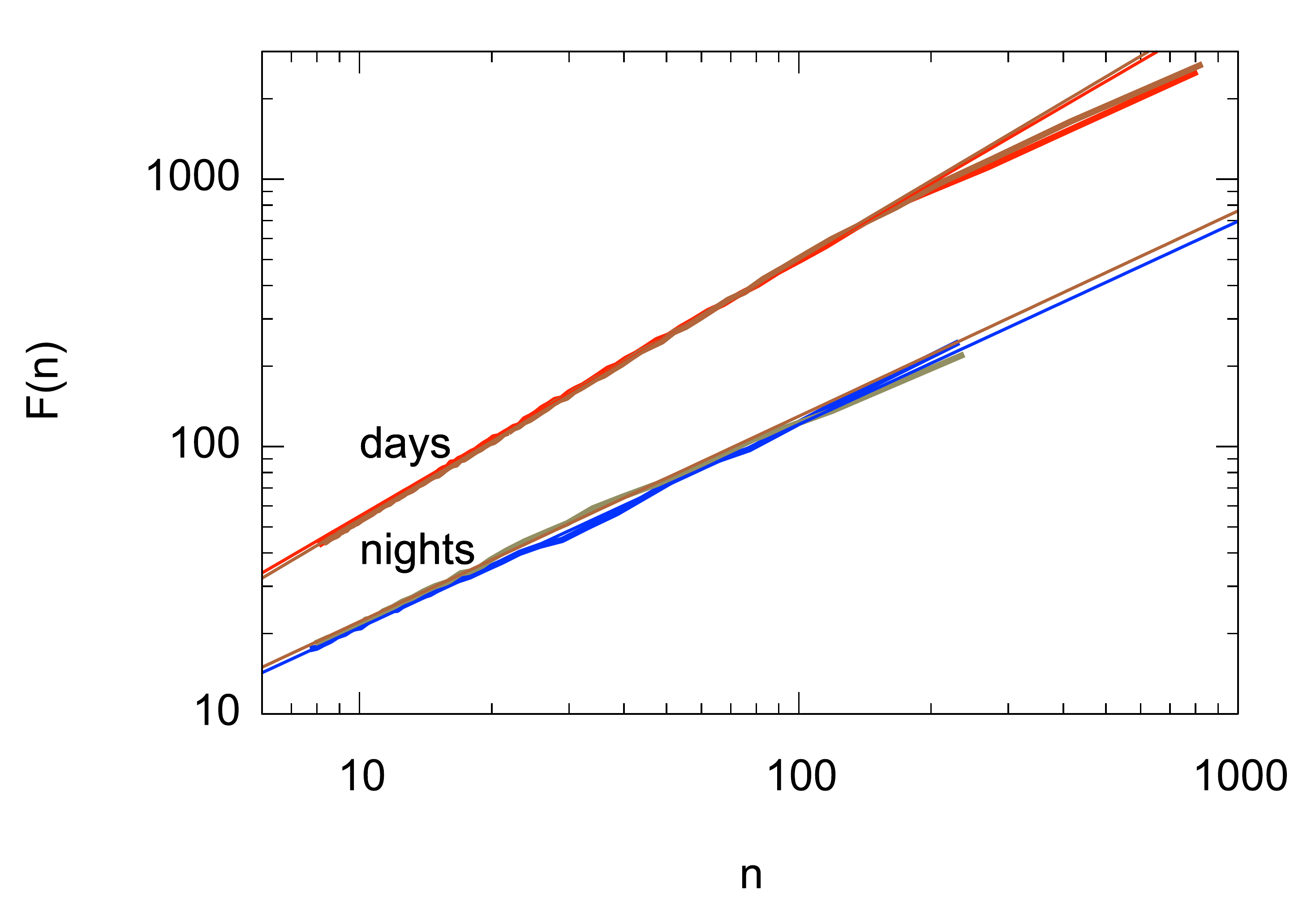}
\caption{\label{h1}DFA measure $F(n)$ plotted against the box size $n$. The red (brown) lines represent day-time data and blue (green) lines are characteristic for nocturnal actigraphy recordings. The $F(n)$ curves are averaged over 18 individuals and over 6 days. The linear fits have been performed in the ranges $8-150$ for days and $7-60$ for nights. Results for healthy sleepers: $H_d=0.9568\pm0.0019$ for day activity and $H_n=0.7599\pm 0.0095$ for nights. Analogous measures for individuals deprived from sleep: $H_d=0.9760\pm0.0015$, $H_n=0.7681\pm 0.0065$.}
\end{figure}

In order to quantify time-correlations in recorded time series, we have used the DFA method \cite{Peng,Talkner}, frequently applied in analysis of physiological data\cite{Holloway,Rodriguez,Danka,Jacek,Sahin}. The original time series 
$X(t),\;t=1,...,N$ have been integrated $Z(k)=\sum^k_{t=1}[X(t)-\overline{X}]$, where $\overline{X}=\sum_{t=1}^N X(t)/N$ is the data mean and $k$ is an integer number $k\in \{1,...,N\}$. Subsequently, the series $Z(k)$ have been divided into boxes of equal length $n$ and, by linear regression analysis, the linear trend of data in each window has been obtained. In the next step, the series $Z(k)$ have been detrended by subtracting the local trend $Z_n(k)$ from the data and the root-mean-square fluctuation has been calculated $F(n)=\sqrt{\frac{1}{N}\sum_{k=1}^N[Z(k)-Z_n(k)]^2}$. The procedure has been repeated for different box sizes ($n=8-840$ min for days and $n=7-220$ min for nights, respectively) and the relationship $F(n)\propto n^{H}$ has been established with $H$ being the Hurst exponent. The complexity measure $H$ provides information \cite{Peng,Talkner} about the memory of investigated dynamic system: when $H=1/2$, the variations in the time series are random and uncorrelated with each other.
For $0<H<1/2$, variations are likely to be antipersistent, i.e., increases of recorded values will be followed by subsequent decreases, whereas for $1/2<H<1$ persistent behaviour of increment variations is most likely observed with increases (decreases) followed by increases (decreases), respectively.

\begin{figure}[h!]
\includegraphics[angle=0,scale=0.5]{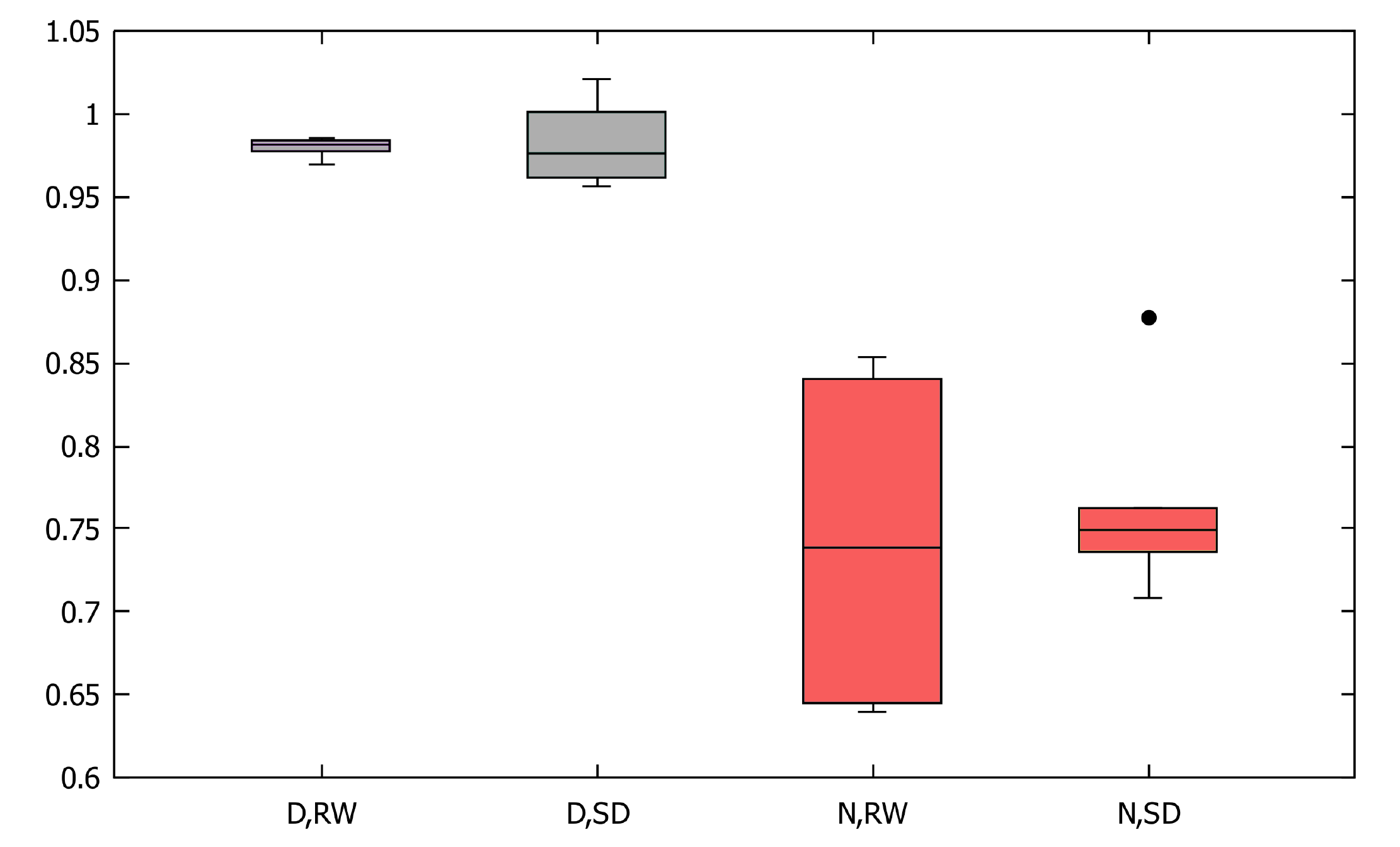}
\caption{Boxplot illustrating variability of the DFA exponents for RW and SD groups over 6 consecutive days (the lines within boxes are medians; the upper and lower box edges represent values of the third and the first quartiles, respectively). The daytime Hurst exponents are labelled: D,RW and D,SD; the exponents for night: N,RW and N,SD.
The subjects have lower value of the mean DFA exponent for nocturnal activity. No difference between the groups is found.}
\label{h2}
\end{figure}

Figures \ref{h1}-\ref{h2} illustrate results of the detrended fluctuation analysis (DFA) performed on the data. 
As can be inferred from Fig. \ref{h1}, the $H$ exponent derived for the mean value of 18 subjects analyzed over  6 days of recordings is greater than 
 the value $1/2$, typical for time-series representing an uncorrelated white noise. All extracted $H$ values are close to 1, thus signaling persistent time-correlations in subsequent intervals. This observation  stays in line
 with the findings based on the other methods:
The exponent $d\approx 1$, as obtained from FF and AF measures or derived from the power spectrum \cite{Ochab}, implies the expected value $H=(d+1)/2\approx 1$.
Moreover, results of analysis presented in Fig.\ref{h1} suggest a more pronounced difference between the groups of healthy sleepers and individuals deprived of sleep in day-activity recordings: $H_d=0.9568\pm 0.0019$ versus $H_d=0.9760 \pm 0.0015$ for RW and SD, respectively. Note, that reported values $\pm 0.0019$  and  $\pm 0.0015$ refer to a standard error (SE) of the fit.

Further analysis of day-time recordings for 18 subjects shows relatively low variability of the evaluated $H_d$ parameters: In Figure \ref{h2} 
each box represents results of DFA performed over all subjects yielding one exponent for each separate day.
In contrast to Fig.\ref{h1},  the comparison of daily exponents ($F(n)$ still being averaged over subjects) excludes clear differentiation
between the groups (no evidence of statistically significant difference at the $95\%$ level of the two-sided two-sample $t$-test, p-value $0.850$).

Unlike the derived $H_d$ values, the complexity exponent $H_n$ obtained for nocturnal recordings, averaged over all subjects and over 6 subsequent days, reflects lower values and larger variation: $H_n=0.7681 \pm 0.0065$ (SE) for SD and $H_n=0.7599 \pm 0.0095$ (SE) for RW,
which already precludes discrimination between the two groups (see Fig.\ref{h1}).
The high variability of the daily derived $H_n$ exponents (see Fig.\ref{h2}) obscures those findings even more, and significance testing indicates no clear difference between SD and RW groups (test as above, p-value $0.684$).

\subsection{Survival probability in active and inactive states}
\label{sec:survival}

To assess the statistics of rare events in tails of waiting time PDF's, as the main measure of the discussed phenomena, we have constructed the (complementary) cumulative distribution {$C(a)$} of durations {$a$}
\begin{eqnarray}
C(a) \equiv Prob(T \ge a) = \int_a^{\infty}P_T(\tau)d\tau,
\end{eqnarray}
which represents the survival probability for the system to stay in a given state up to the time $a$.  For a renewal process with the counting function $N(t)$ indicating the effective number of events before or at instant of time $t$: $N(t)\equiv max\{k|t_k\leq t, k=0,1,2,...\}$, the survival function can be identified as $C(a)=Prob(N(a)=0)$.
For a stationary time series following a Poisson renewal process the survival probability $C(a)$ is expected to have a characteristic scale (a relaxation time {$\tau_{rel}$}) related to the probability $Prob(N(t)=0)=e^{-t/\tau_{rel}}=e^{-rt}$. In contrast, for non-stationary complex systems with inhomogeneous distributions of decay times
  $\tau_{rel}$, the resulting complementary distribution $C(a)$ may be scale invariant \cite{Bickel,Palva,Ochab}. 

To further asses this point, we have assumed that the human subjects can be in two states (active or resting) with some dwell time distributions characterizing duration of active and resting phases. We have defined a subject to be in the active state if the experimental time series $X(t)>85$, and to be at rest otherwise. The choice of an appropriate threshold is argued in Supplementary Material to \cite{Ochab}.
The extracted estimates of cumulative distributions have been fitted with two formulae: a power-law
$C(a) = a^{-\gamma}$
for rest periods and a stretched exponential
$C(a) = \exp\left (-\alpha a^\beta \right )$
for activity periods. The fitting has been performed using log-log or log-linear data, respectively, in order to account for the tails in the distributions.
The fitted parameters $\alpha$, $\beta$, and $\gamma$ have been then compared for the RW and SD conditions.
\begin{figure}[htb]
 \includegraphics[width=0.49\columnwidth]{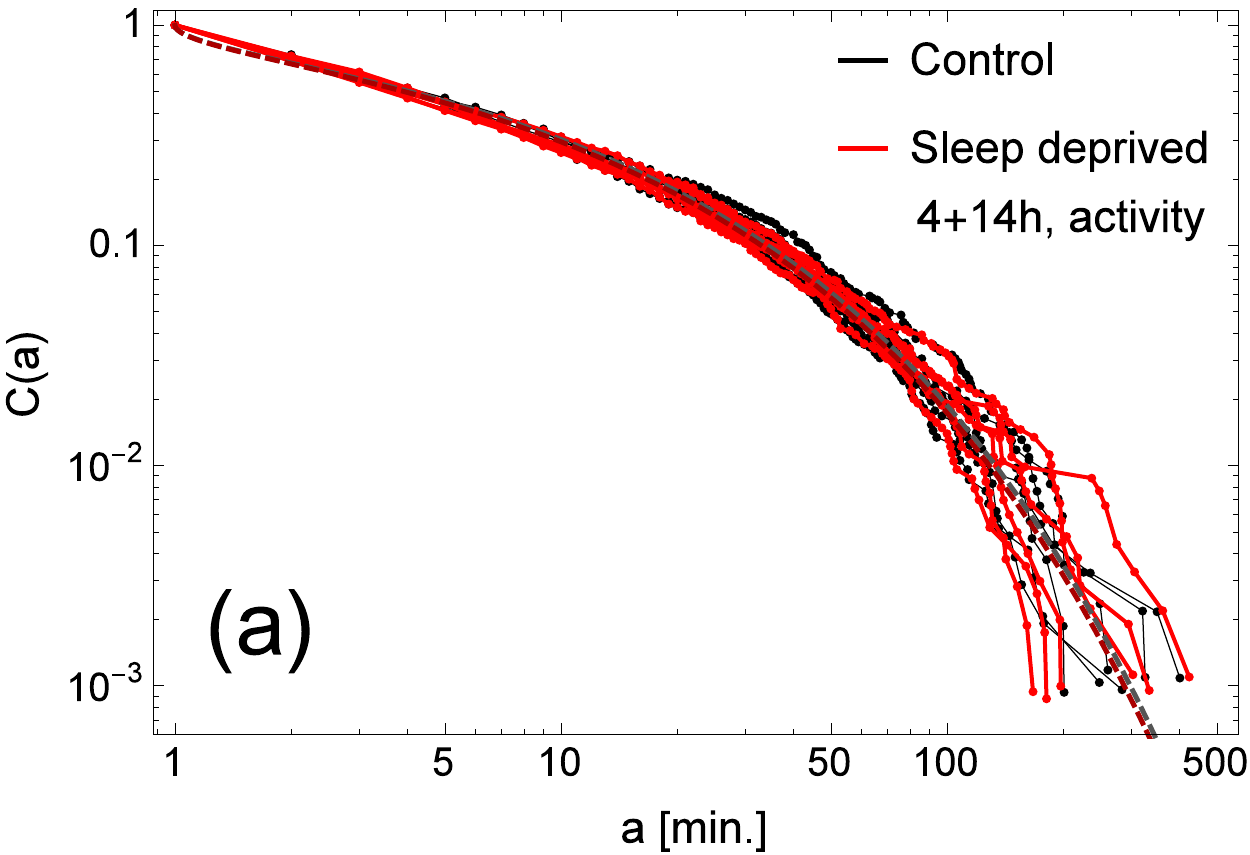}
 \includegraphics[width=0.49\columnwidth]{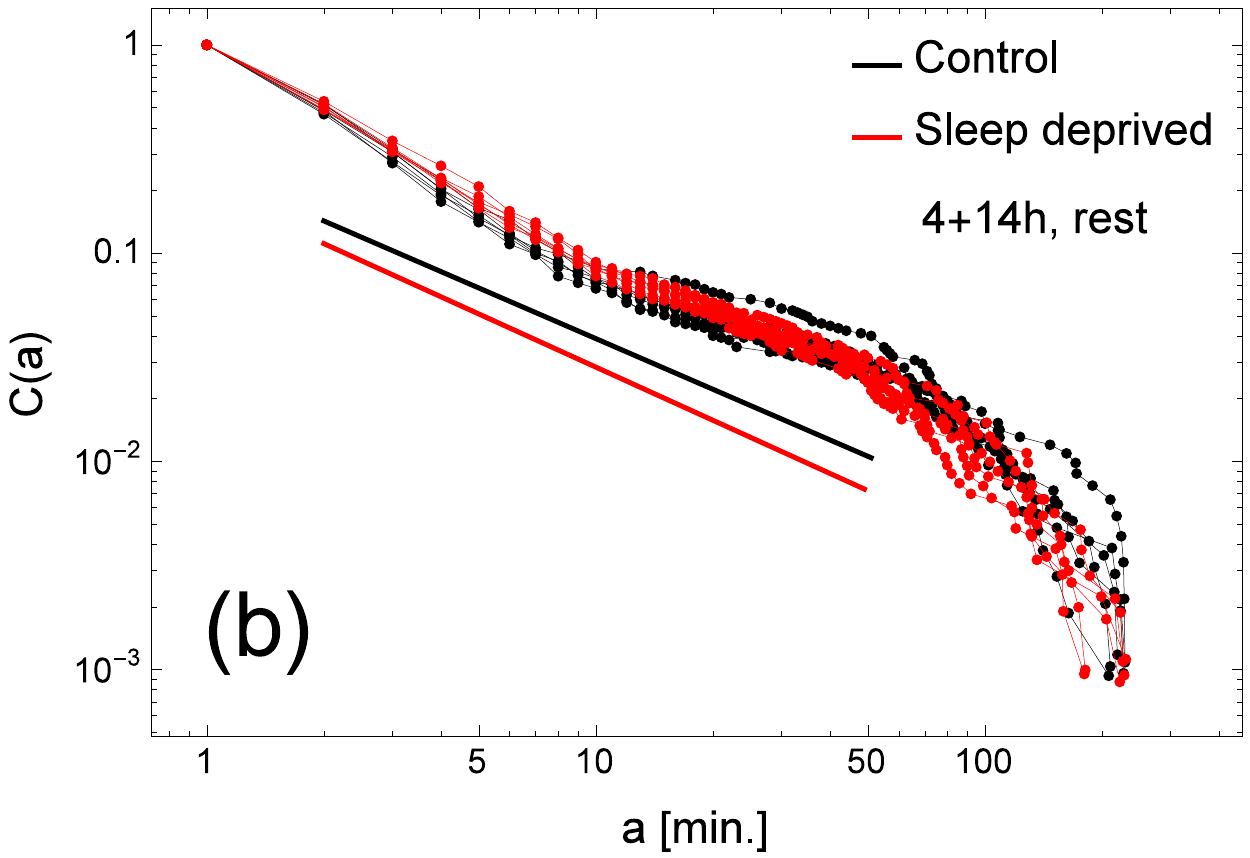}
\caption{\label{c1} Complementary cumulative distribution of (a) activity and (b) rest periods for the control (black) and sleep deprived (red) group. Each curve corresponds to a distribution gathered from all the valid subjects on a single day (consisting of 4h sleep followed by 14h daytime activity). The activity and rest periods were defined as the times where ZCM was respectively above or below 85. Dashed lines in (a) show fitted stretched exponentials; solid lines in (b) show the average (over 6 days) fitted slope, and extend across the maximal range of fitting.}
\end{figure}

\begin{figure}[htb]
\includegraphics[width=0.69\columnwidth]{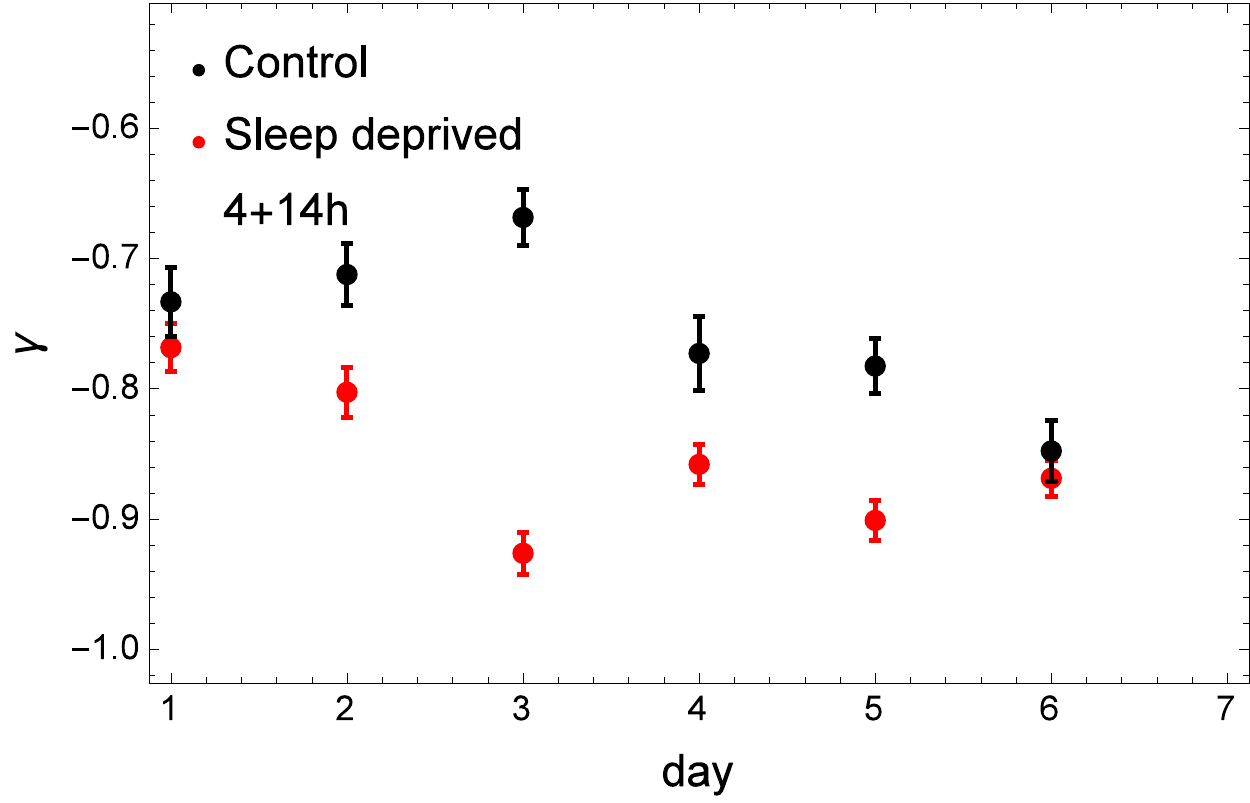}
\caption{\label{c3} The power-law exponents $\gamma$ fitted to the complementary cumulative distributions from Fig. \ref{c1}b for each consecutive day (4h sleep followed by 14h daytime activity). The error bars indicate the standard error of the fit.}
\end{figure}

Each cumulative distribution, i.e., each individual line shown in Fig.~\ref{c1} has been constructed from rest/activity periods collected over a single day from actigraph recordings of all 18 participants under RW or SD conditions. 
Consequently, as shown in Fig.~\ref{c3}, for each day one data point characterizing the distribution of the whole group in RW (control) or SD condition has been obtained.

The parameters characterizing a single cumulative distribution of rest/activity periods have been further averaged over the whole week.
In order to compare RW and SD groups, such means of parameters fitted for several consecutive days have been compared
with the two-tailed $t$-tests performed at 95\% confidence level,
preceded by a set of tests for equal variances, as explained elsewhere \cite{Ochab}. In the case of activity periods no clear difference between the RW and SD individuals has been observed. The overall fit (averaged over days and individuals) yields for the RW sample $\bar{\alpha}=0.31\pm0.04$ and $\bar{\beta}=0.49\pm0.03$. Similar values for the SD mode are not significantly different ($p>0.05$; two-tailed $t$-test).

Unlike these findings, the cumulative distributions $C(a)$ of rest period durations, averaged over all participants indicate a significant difference in behavioural motifs between the control group and sleep deprived individuals.
Specifically, the means $\bar{\gamma}$ of exponents for RW and SD subjects are different at 5\% level, with p-value $0.012$ (for 4-hour sleep followed by 14-hour daytime activity); additional leave-one out cross-validation was performed with the same conclusion. As mentioned before in Sec.~\ref{sec:FF}, the reverse scheme of 14-hour daytime activity followed by 4-hour sleep (the length of time series stays as before, but different parts of data are cut away) was checked as well. In this instance, the means $\bar{\gamma}$ do not differ at 5\% level, with p-value $0.065$. 
The higher (average) coefficient $\bar{\gamma}= 0.85\pm 0.03$ extracted for sleep-deficient individuals emphasizes the fact that the pattern of their resting times consists of more short periods and, respectively, fewer longer inactivity time intervals than in the control group.

\section{Discussion and conclusions}
Vulnerability to chronic sleep deprivation is a complex phenomenon in which performance, subjective and neural measures indicate distinct features, possibly related to chronotype, somnotype and gender \cite{Oginska,Sun,Holloway,NakamuraPLOS,Matuzaki}. Although not used routinely as a measurement of choice for evaluating the severity of locomotor disorders related to sleep deprivation in patients, the actigraphy records are considered by many researchers and medical doctors a valuable tool for demonstrating behavioural disturbances related to sleep deficit.
Information detected in actigraph data provide measures for sleep and wake motifs in healthy individuals and in people with certain sleep problems and can be used to establish links between duration of sleep and fatigue and performance effectiveness \cite{Matuzaki}. In particular, it seems that a strategy of assessment based on use of actigraphy recordings may serve as an objective tool to discriminate behavioural patterns associated with shortness of sleep and insomnia \cite{Holloway,Sun,Kim,Nakamura}.

In this paper properties of actigraph time-records have been analysed by use of multiple complexity measures. 
Presented model of a non-uniform random point process had successfully explained the main properties of analysed time-series indicating clustering of event counts as measured by Fano and Allan factors. Provided estimates allow to detect correlations in inter-event periods typical for a fractal stochastic process \cite{Anteneodo,Klafter,Mainardi,Eden}. Both factors increase with the counting time window as $T^d$ suggesting variability of the response (activity counts) characterized by the autocorrelation function: The linear growth in $T$ of the event count variance is characteristic of a diffusive process. In contrast, the FF value increasing as a power function of the counting window indicates that fluctuations in the rate are observable on many time scales with a self-similar exponent $d$. The  Fano factors derived in our study are always larger than 1, documenting that the rate of events in the data is an inhomogeneous process with aggregated event-counts. This  super-Poissonian behaviour is observed for  both\textemdash nocturnal and day-time recordings. Similarly, the Allan factor which allows to detect the average variation in the pattern of adjacent counts, also points to the time correlations in analysed series. Nevertheless, variations of extracted FF and AF coefficients in both groups of examined subjects exclude drawing inference in hypothesis testing and do not allow to discriminate between the degree of correlation and patterns of bursting in the counts
of events observed for the control (RW) or sleep-deprived (SD) individuals. 
In parallel to these findings, the complexity scores based on DFA analysis stay in conformity with the observed self-affinity of activity signals. However, while series of nocturnal recordings display a higher median of the Hurst exponent for the SD group with respect to the RW subjects, applied significance tests show no clear difference between the control group and the group with sleep deficiency. That observation limits practical use of DFA analysis in the identification of reduced sleep dynamics, in contrast to results on sleep impaired subjects diagnosed with acute insomnia \cite{Holloway}.


To further identify the best mathematical model behind measured activity signals, they can be tested for stationarity and ergodicity \cite{Magdziarz}. Verification of stationarity of the time series can be approached by quantile lines test, followed by checking ergodic behaviour \cite{Burnecki} to validate anomalous diffusion models of the data. These can be complemented by tests based on power spectra of FARIMA model \cite{Preuss}. While so far we have based our conclusions on fairly simple properties of the experimental data, there are indications of their non-stationarity. These issues are a starting point for our further research.

Altogether, our investigations of long-range correlations present in the analysed time-series provide evidence that the best measure of discrimination between healthy sleepers and sleep-deficient group is obtained by examination of the cumulative (survival) distribution of resting periods. Even though period durations are limited by the time resolution of an actigraph and the night length,
the survival distribution in a resting state exhibits approximately a power-law scaling which is robust over two decades.
Significantly higher values of the exponent $\gamma$ for sleep-deprived subjects signal less heavy tails of waiting-time distributions in an immobile (resting) state than in an analogous distribution for the control group
and can be associated with restlessness/inquietude and increased variability (burstiness) of activity in recorded time series.

Consequently, those alterations of locomotor pattern can be informative about mood/behaviour disturbances related to sleep deficiency and possibly, used as a valuable diagnostic fingerprint discriminating between healthy and depressed/disordered individuals.

\section*{Acknowledgments}
This work was supported in part by National Science Center (ncn.gov.pl) grants No. DEC-2011/02/A/ST1/00119 (JO) and DEC-2014/13/B/ST2/020140 (EGN), and the Polish Ministry of Science and Higher Education grant No. 7150/E-338/M/2014 (KO).
The project communicated to Unsolved Problems of Noise 2015, July 14-17, Barcelona, Spain.

\end{document}